\documentclass[10pt, a4paper]{article}
\usepackage{lrec}
\usepackage{graphicx}
\usepackage{tabularx}
\usepackage{soul}
\usepackage{etoolbox,siunitx}
\usepackage{textcomp}
\renewcommand{\arraystretch}{1.08}
\usepackage{soul}
\usepackage{array}
\usepackage{textgreek}
\usepackage{tabularx}
\usepackage{booktabs}
\usepackage{romannum}
\usepackage{textcomp}
\usepackage{subfigure} 
\robustify\bfseries

\usepackage{epstopdf}
\usepackage[latin1]{inputenc}
\usepackage{amsmath}
\usepackage{hyperref}
\usepackage{xstring}
\usepackage{multirow}

\usepackage{cellspace}
\newcolumntype{L}[1]{>{\raggedright\let\newline\\\arraybackslash\hspace{0pt}}m{#1}}
\newcolumntype{C}[1]{>{\centering\let\newline\\\arraybackslash\hspace{0pt}}m{#1}}
\newcolumntype{R}[1]{>{\raggedleft\let\newline\\\arraybackslash\hspace{0pt}}m{#1}}

\newcolumntype{P}[1]{>{\centering\arraybackslash}p{#1}}

\title{AccentDB: A Database of Non-Native English Accents to\\ Assist Neural Speech Recognition}

\name{Afroz Ahamad\textsuperscript{{*}\textdagger}\thanks{* denotes equal contribution from the authors.},
\,Ankit Anand\textsuperscript{*}\thanks{\textdagger \,currently at Google.}, \,Pranesh Bhargava }

\address{BITS Pilani, India \\
         \{afrozsahamad, ankit0905anand\}@gmail.com\\
         pranesh@hyderabad.bits-pilani.ac.in\\}

\abstract{Modern Automatic Speech Recognition (ASR) technology has evolved to identify the speech spoken by native speakers of a language very well. However, identification of the speech spoken by non-native speakers continues to be a major challenge for it. In this work, we first spell out the key requirements for creating a well-curated database of speech samples in non-native accents for training and testing robust ASR systems. We then introduce AccentDB, one such database that contains samples of 4 Indian-English accents collected by us, and a compilation of samples from 4 native-English, and a metropolitan Indian-English accent. We also present an analysis on separability of the collected accent data. Further, we present several accent classification models and evaluate them thoroughly against human-labelled accent classes. We test the generalization of our classifier models in a variety of setups of seen and unseen data. Finally, we introduce the task of accent neutralization of non-native accents to native accents using autoencoder models with task-specific architectures. Thus, our work aims to aid ASR systems at every stage of development with a database for training, classification models for feature augmentation, and neutralization systems for acoustic transformations of non-native accents of English.
\\ \newline  }

\begin{document}
\maketitleabstract

\section{Introduction}

In Sociolinguistics, accent is a manner of pronouncing a language. Anyone who speaks a language, does so in an accent. The way the native speakers of a language speak that language defines the standard pronunciation, and is generally considered to be the standard or reference accent for that language. When the non-native speakers of a language speak that language, say an Indian person speaking English, the phonological requirement of the non-native language, in this case English, interacts with the phonological knowledge of their first language, say Hindi. This influences their manner of speaking, giving rise to what is considered as the non-native accent.

Accents \textit{per se} are interesting because they refer to a wide variety of social issues such as the acceptance of speakers into a community, indication of class in society, and linguistic issues such as those pertaining to the phonology of languages. This in itself warrants a better understanding of accents. However, there is another fundamental reason for studying accents. Speakers always have a manner of speaking and the speech always has accent. Since spoken communication is an important form of communication, studying accents becomes important to design technologies built to interact with human speech.
\subsection{Indian Accents in English}
Internet has led to English language becoming the lingua\-franca for conveying information about science, culture, sports and society in the world. The continued advancements in technologies supporting speech, in the form of audio and video media, has led to an increase in the usage of spoken English on the web. Since these speakers come from various different linguistic backgrounds, English language happens to be spoken in many different accents across the world. 

English has become an important language of communication among the younger generation of India because of its status as the language of formal education. A large number of young Indians is bilingual, i.e. they speak one of the 22 Indian languages as their first language, alongside English. An implication of this is that when speaking English, the intervention from the phonology of their first language, e.g. Malayalam, gives rise to an accent in the speech of Indian speakers of English. This accent is generally very distinct and is readily identifiable, for example, as the Malayalam English accent, the Telugu English accent, the Bangla English accent, etc. 

Interestingly, this younger generation of India is also a large and growing group of users of speech-based technology through hand-held devices and voice assistants. These voice assistants have become very good at identifying English spoken in a native accent. However, non-native accented speech continues to be a challenge for them. If the automatic speech recognition (ASR) systems of the voice assistants have \textit{apriori} knowledge that the speaker is going to speak with a certain accent, the voice assistant may be primed to listen to certain features in the voice, which would lead to a greater performance accuracy. For the success of this technology, it becomes pertinent then to identify and process accents, apart from the semantic content of the speech. Due to the large number of speakers, and vast varieties of accent, English spoken within India is an excellent resource for creating and testing technology whose success is contingent on detecting, identifying and understanding the native and non-native accents.

\section{The Database}
A key requirement for developing speech-based technology is the access to a well-curated database of speech samples. Some of the widely used datasets for specific ASR tasks are very well labelled, either manually or through automation. For example, Google AudioSet \cite{audioset} is a massive dataset for audio event detection, that includes more than 2 million manually-labelled 10-second sound clips belonging to over 600 classes. Similarly, VoxCeleb \cite{voxceleb} is a speaker identification dataset which contains audio clips extracted from interviews of celebrities. 

In this section, we first establish certain key requirements for constructing an accent database that could be well-suited for ASR tasks. Then we survey a few existing accent datasets. Further, we discuss our approach and setup for collecting our database, AccentDB\footnote{\url{https://accentdb.github.io/}}. Finally, we present an analysis of the distribution of speech samples that constitute AccentDB.

\subsection{Key Requirements}
The following are some of the key requirements for an accent database suitable for ASR systems.

\romannum{1}. \textbf{Variety of Speakers:} In order to represent the speaker differences, the database should ideally contain spoken material from a wide range of speakers.

\romannum{2}. \textbf{Words vs. Sentences:} The pronunciation patterns for words spoken in isolation are different from when they appear in connected speech, due to the suprasegmental phenomena such as elision and assimilation \cite{Ladefoged}. Therefore, for the purposes pertaining to the processing of spoken sentences, the database should contain sentence-length material.

\romannum{3}. \textbf{Uniformity of Content:} For the sake of isolating and identifying accents, it is necessary to have uniformity in the speech material across speakers. One way to address this is to have all the speakers speak the same sentences, preferably at the same speed. A related requirement is for the speech material to be phonetically balanced, so that no specific phonemes get over-represented in the database.

\romannum{4}. \textbf{Semantic Requirement:} If the sentences are meaningful, it avoids semantic factors affecting the pronunciation of the sentences.

\subsection{Existing Accent Databases}

Various attempts have been made in the past at creating accent focused speech databases with varied data sources, speakers, accents and corpora.  \newcite{carlos-isabel-antonio} created a word database with 20 speakers for each accent from a total of 6 countries. They used a small corpus of around 200 isolated English words spoken twice in a row by each speaker. \newcite{deep-learning-for-video-games} presented a collection of British and American accents in the form of utterances from non-playable characters of the video game, "Dragon Age: Origins (BioWare 2009)", with manual labelling of the accents done by three individuals.
\\
Two of the most popular datasets used for accent-related tasks are: the Foreign Accented English (FAE) corpus \cite{foreign-accented-english}, and the Speech Accent Archive \cite{please-call-stella}. FAE data comprises 4925 telephonic utterances by native English speakers of 22 different languages. The subjects spoke about themselves for 20 seconds and the recordings were rated on a 4-point scale to determine the strength of accent. 
The Speech Accent Archive is a crowd-sourced collection of speech recordings of readings of a passage (colloquially referred to as \textit{"Please call Stella.")} in English. Information about speakers' demographic and linguistic background is publicly available \footnote{\href{http://accent.gmu.edu/browse_language.php}{Speech Accent Archive, George Mason University.}}. The passage has been spoken by more than 2000 speakers covering over 100 accents and 30 languages, but a significant number of samples are not tagged with the correct accent. This is because the database is crowd sourced, and there is no independent supervision on the accent label that is assigned to a recorded audio sample. For instance, a speaker whose first language is Bengali/Bangla, might mark his samples as belonging to the Bangla accent, even if his Bangla accent is neutralized after living in the UK for many years. Another drawback of using such crowd sourcing approaches for collection of accent data is that neither the recording environment, nor the recording hardware are consistent across speakers. This leads to the introduction of significant noise in samples. The lack of correct label for each sample adds to the difficulty of using any supervised learning algorithm for speech recognition tasks. 

\begin{table}[t]
\centering
\begin{tabular}{l}
\midrule
\textit{The birch canoe slid on the smooth planks.}\\
\textit{Glue the sheet to the dark blue background.}\\
\textit{It's easy to tell the depth of a well.}\\
\textit{These days a chicken leg is a rare dish.}\\
\textit{Rice is often served in round bowls.}\\
\midrule
\end{tabular}
\caption{First 5 sentences of the Harvard Sentences dataset.}
\label{tab:thebirchcanoesentences}
\end{table}

\renewcommand{\arraystretch}{1}
\begin{table*}[ht]
\centering
\begin{tabular*}{0.805\textwidth}{llclc}
\toprule
 & \textbf{Accent} & \textbf{Number of Samples} & \multicolumn{1}{p{2cm}}{\centering \hspace{0.1cm} \textbf{Duration}} & \textbf{Number of Speakers} \\ \midrule
 \multirow{5}{*}{\textbf{AccentDB}} 
&Bangla & \tablenum[table-format=5]{1528}  & \tablenum[table-format=2]{2}h \tablenum[table-format=2]{13}min & \tablenum[table-format=2]{2} \\
&Malayalam & \tablenum[table-format=5]{2393} & \tablenum[table-format=2]{3}h \tablenum[table-format=2]{32}min & \tablenum[table-format=2]{3} \\
&Odiya & \tablenum[table-format=5]{748} & \tablenum[table-format=2]{1}h \tablenum[table-format=2]{11}min & \tablenum[table-format=2]{1} \\
&Telugu & \tablenum[table-format=5]{1515} & \tablenum[table-format=2]{2}h \tablenum[table-format=2]{10}min & \tablenum[table-format=2]{2} \\[0.5mm]  \cmidrule{2-5} 
& {\textbf{Total}} & \bfseries \tablenum[table-format=5]{6184} & \bfseries \tablenum[table-format=2]{9}h \tablenum[table-format=2]{06}min & \bfseries \tablenum[table-format=2]{8} \\ \midrule

\multirow{6}{*}{\textbf{Amazon Polly}}
&American & \tablenum[table-format=5]{5760} & \tablenum[table-format=2]{5}h \tablenum[table-format=2]{44}min & \tablenum[table-format=2]{8} \\
&Australian & \tablenum[table-format=5]{1440} & \tablenum[table-format=2]{1}h \tablenum[table-format=2]{21}min & \tablenum[table-format=2]{2} \\
&British & \tablenum[table-format=5]{1440} & \tablenum[table-format=2]{1}h \tablenum[table-format=2]{26}min & \tablenum[table-format=2]{2} \\
&Indian & \tablenum[table-format=5]{1440} & \tablenum[table-format=2]{1}h \tablenum[table-format=2]{29}min & \tablenum[table-format=2]{2} \\
&Welsh & \tablenum[table-format=5]{720} & \tablenum[table-format=2]{0}h \tablenum[table-format=2]{43}min & \tablenum[table-format=2]{1} \\[0.5mm]  \cmidrule{2-5} 
& {\textbf{Total}} & \bfseries \tablenum[table-format=5]{10800} & \bfseries \tablenum[table-format=2]{10}h \tablenum[table-format=2]{43}min & \bfseries \tablenum[table-format=2]{15} \\ \midrule
\multicolumn{2}{l}{\textbf{\, Total}} & \bfseries \tablenum[table-format=5]{16984} & \bfseries \tablenum[table-format=2]{19}h \tablenum[table-format=2]{49}min & \bfseries \tablenum[table-format=2]{23} \\ \bottomrule
\end{tabular*}
\caption{The details of total \SI{9}{} accents: \SI{4}{} collected by the authors and \SI{5}{} compiled using Amazon Polly.}\label{tab:our_dataset}
\end{table*}
\renewcommand{\arraystretch}{1.08}

The CMU Festvox Project has a dataset titled CMU-Arctic \cite{cmu-arctic} which contains speech samples in native English accents. In CMU-Indic, another dataset in the Festvox project, the content across the samples is not uniform as they are spoken not in one language with different accents, rather in different languages altogether. The samples here incorporate certain manifestations of an accent as well, as is evident from samples in any Indian language such as Gujarati, but the task of accent classification now entails modelling two attributes - the difference in utterances and the accent itself.\\

\subsection{Introducing AccentDB}
\begin{table}[h]
\small
\centering
\begin{tabular*}{0.4805\textwidth}{clccc}
\toprule
{Speaker} & \multicolumn{1}{c}{{Native}} & {Age of} & \multicolumn{1}{c}{Highest} & {English} \\
{Code} & \multicolumn{1}{c}{{Language}} & {Speaker} & \multicolumn{1}{c}{Qualification} & {Usage} \\ \midrule
Ban-1 & Bangla & \tablenum[table-format=2]{24} & Masters & \tablenum[table-format=2]{19}yrs  \\ 
Ban-2 & Bangla & \tablenum[table-format=2]{25} & Masters &  \tablenum[table-format=2]{21}yrs \\ 
Mal-1 & Malayalam & \tablenum[table-format=2]{25} & Masters & \tablenum[table-format=2]{20}yrs  \\ 
Mal-2 & Malayalam & \tablenum[table-format=2]{25}  & Masters  & \tablenum[table-format=2]{20}yrs  \\ 
Mal-3 & Malayalam & \tablenum[table-format=2]{26} & Masters & \tablenum[table-format=2]{21}yrs  \\ 
Odi-1 & Odiya & \tablenum[table-format=2]{33} & Ph.D. &  \tablenum[table-format=2]{28}yrs \\ 
Tel-1 & Telugu & \tablenum[table-format=2]{26} & Masters &  \tablenum[table-format=2]{21}yrs \\ 
Tel-2 & Telugu & \tablenum[table-format=2]{32} & Ph.D. &  \tablenum[table-format=2]{17}yrs \\ \bottomrule
\end{tabular*}
\caption{Demographic details of the speakers of speech samples in AccentDB database.}
\label{tab:demographics}
\end{table}
To fulfill the aforementioned key requirements and to avoid the issues faced by some existing databases, we created a multiple-pair parallel corpus of well structured and labelled data of accents. The database, AccentDB, contains speech recordings in 9 accents, split across 4 non-native accents of Bangla, Malayalam, Odiya and Telugu; 1 metropolitan Indian accent referred as "Indian" and 4 native accents namely American, Australian, British and Welsh. The number of samples, duration of all samples and the number of speakers per accent are listed in Table \ref{tab:our_dataset}.

AccentDB is collected by employing the Harvard Sentences \cite{harvard-sentences} which are phonetically balanced sentences that use specific phonemes at the same frequency as they appear in English language. The sentences in this dataset are neither too short nor too long, making them suitable for proper manifestation of accents in sentence-level speech. Harvard Sentences dataset contains 72 sets, each consisting of 10 sentences. The first five sentences from this dataset are listed in Table \ref{tab:thebirchcanoesentences}.  We ensure that the corpus is also parallel by recording a minimum of the same 25 sets across all 4 of the non-native accents. Additionally, we compile recordings of all the 72 sets across rest of the 5 accents. 


\subsection{Collection of Speech Data}
The data for the \SI{4}{} non-native accents, namely Bangla, Malayalam, Odiya and Telugu, was collected by the authors. For the task of recording speech samples, we recruited volunteers whom we identified to have strong non-native English accents in their daily conversations. Another requirement for these speakers was for them to be the native speakers of at least one Indian language since childhood. The demographics of the speakers can be found in Table \ref{tab:demographics}. 

The data was collected in the form of audio recordings made inside a professionally-designed soundproof booth. The text of the sentences was presented to the participants on a computer screen through a web-app\footnote{\url{http://speech-recorder.herokuapp.com/}} designed specifically for this purpose. The participants were asked to read the text of the sentences aloud. The speech samples were recorded using the following equipment:
\begin{itemize}
    \item Microphone : Audio Technica AT2005USB Cardioid Dynamic Microphone 
    \item Recorder: Tascam DR-05 Linear PCM Recorder
\end{itemize}

Each set was repeated thrice to account for the speech variations in each sentence spoken by the same speaker. 

For the 4 native accents, namely British, Welsh, American and Australian, and the metropolitan Indian accent, we generated speech samples by using Amazon Polly's Text-to-Speech API \footnote{\url{https://aws.amazon.com/polly/}}. The API was used with a special speech synthesis markup formatted file\footnote{\href{http://enigmaeth.github.io/about/HarvardSentences.ssml}{\nolinkurl{HarvardSentences.ssml}}} containing the Harvard Sentences.

\begin{figure*}%
\centering
\subfigure[A][PCA transformation in \SI{3}{}-dims]{%
\label{fig:ex3-a}%
\includegraphics[height=1.6in, width=1.7in]{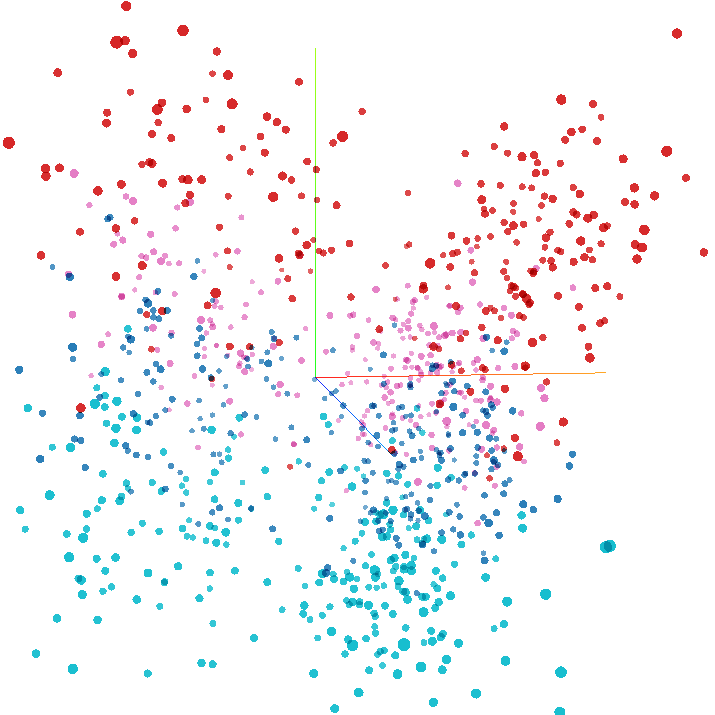}}%
\hspace{20pt}%
\subfigure[][UMAP with \SI{20}{} neighbours]{%
\label{fig:ex3-b}%
\includegraphics[height=1.6in, width=1.7in]{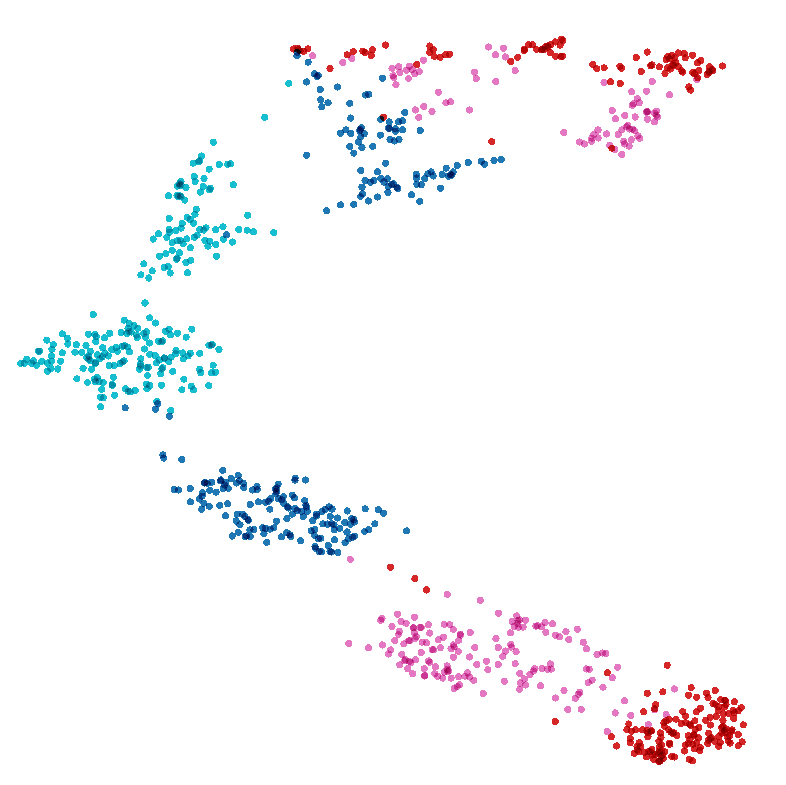}}%
\hspace{20pt}%
\subfigure[][UMAP with \SI{50}{} neighbours]{%
\label{fig:ex3-c}%
\includegraphics[height=1.6in, width=1.7in]{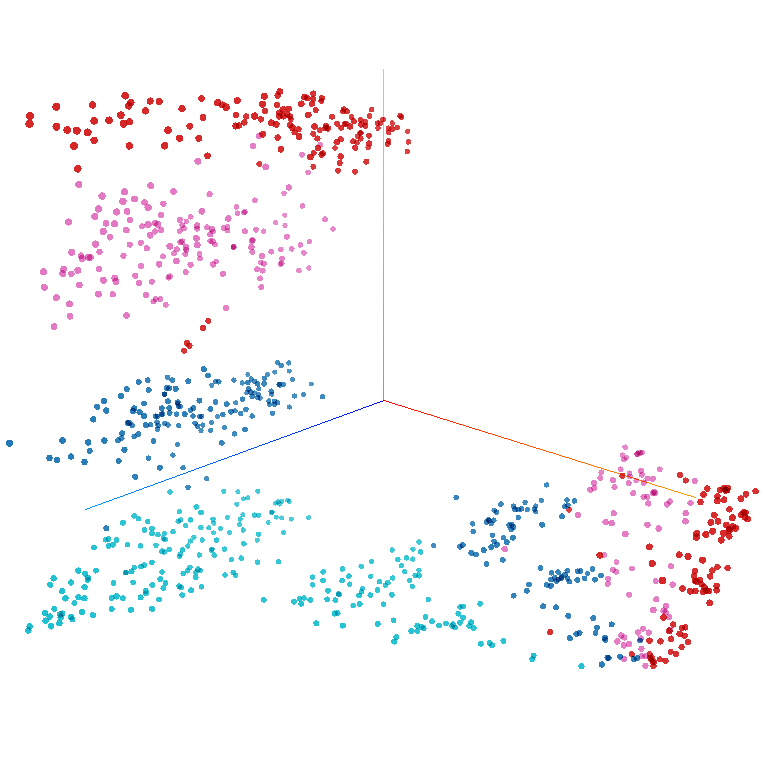}} \\
\subfigure[][t-SNE \textit{(\textrho: 5, \textnu: 1, \textepsilon : 400)}]{%
\label{fig:ex3-d}%
\includegraphics[height=1.6in, width=1.7in]{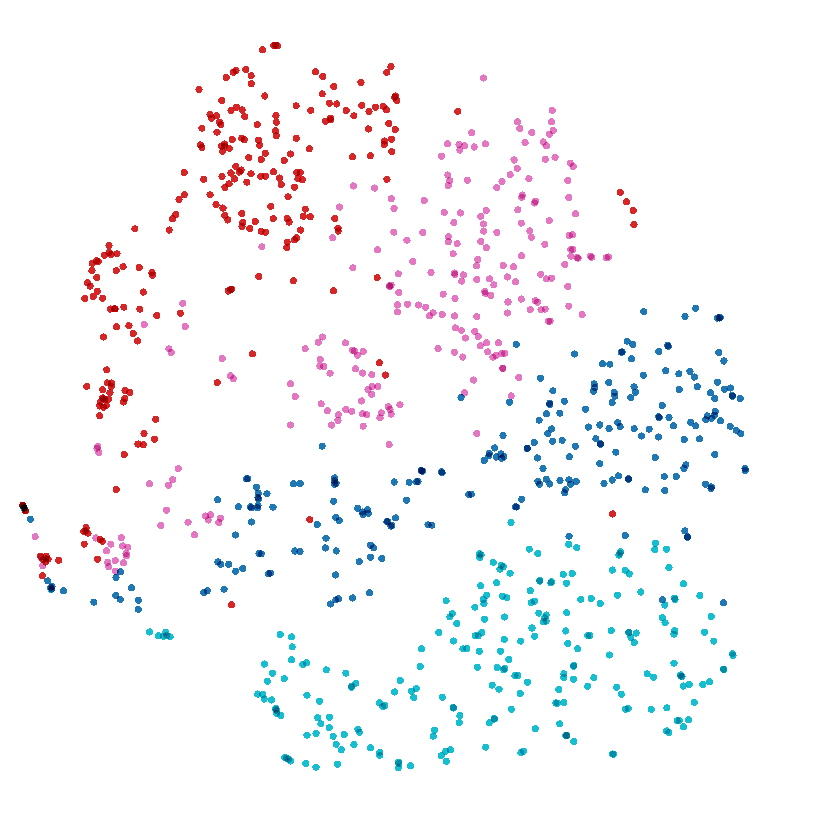}}%
\hspace{20pt}%
\subfigure[][t-SNE \textit{(\textrho: 40, \textnu: 10, \textepsilon : 1600)}]{%
\label{fig:ex3-e}%
\includegraphics[height=1.6in, width=1.7in]{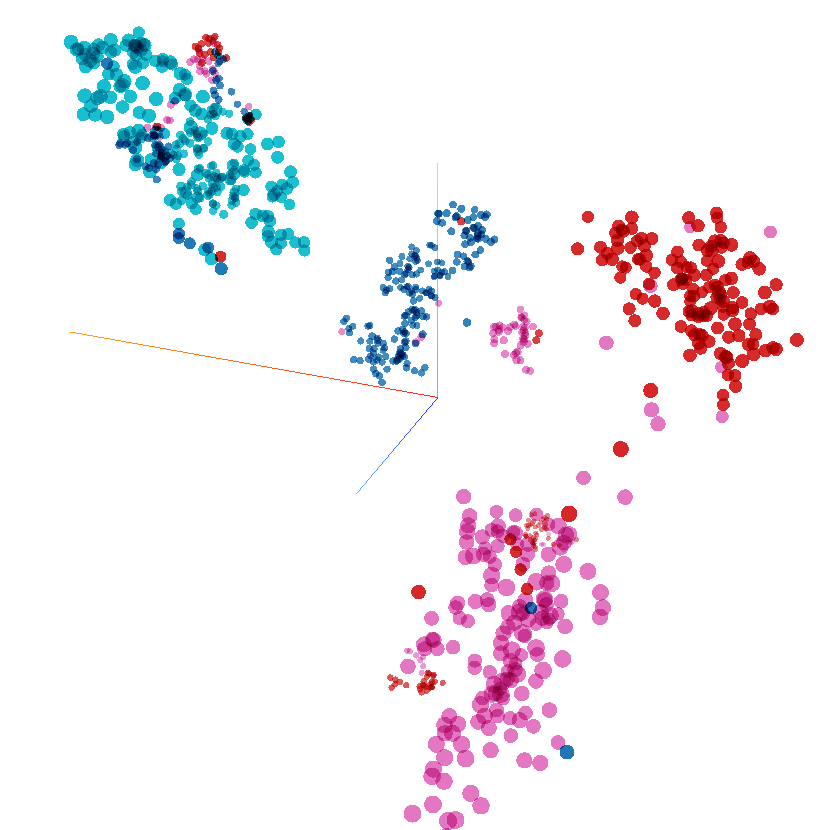}}%
\hspace{20pt}%
\subfigure[][t-SNE \textit{(\textrho: 40, \textnu: 10, \textepsilon : 4000)}]{%
\label{fig:ex3-f}%
\includegraphics[height=1.6in, width=1.7in]{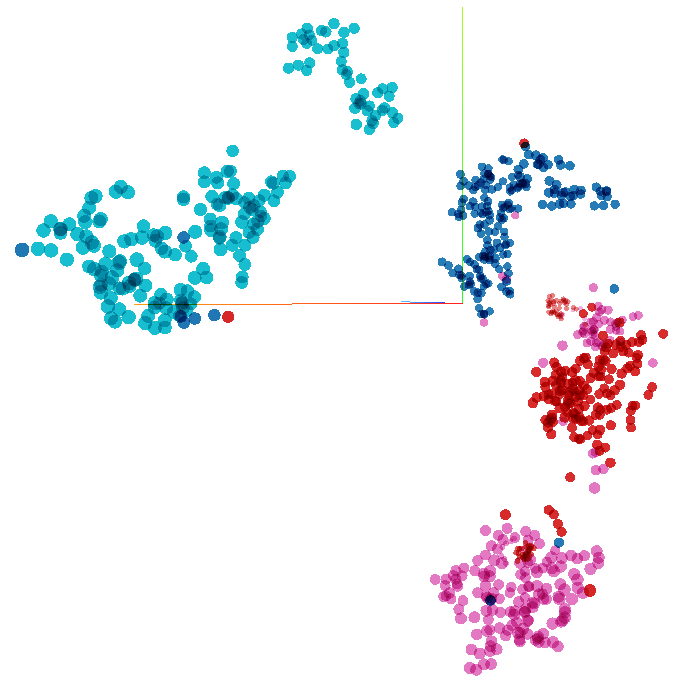}}%
\caption[Projection of MFCC features for 4 non-native Indian Accents.]{Projection of MFCC features for 4 non-native Indian Accents;
\subref{fig:ex3-a}(3D): Principal Component Analysis (PCA) of feature vectors in 3-dimensions; \subref{fig:ex3-b}(2D), \subref{fig:ex3-c}(3D): Uniform Manifold Approximation and Projection (UMAP) with \SI{20}{} and \SI{50}{} neighbours; and \subref{fig:ex3-d}(2D), \subref{fig:ex3-e}(3D), and \subref{fig:ex3-f}(3D): t-distributed Stochastic Neighbor Embedding (t-SNE) projections with different perplexity\textit{(\textrho)}, learning rate\textit{(\textnu)} and number of epochs\textit{(\textepsilon)}.}
\label{fig:ex3}%
\end{figure*}

\subsection{Cleaning and Post-processing }
Any noise or other unwanted events (sneeze, giggle etc.) that were introduced while recording were sliced out using Audacity \cite{audacity} software. The cleaned audio files consisting of more than an hour-long recordings from each speaker were split on a pre-computed silence threshold to make one audio file per sentence. A split was created wherever the energy level was below \SI{1.0}{\%} for a duration of atleast \SI{2} seconds. We then also trimmed silence slices at the beginning and the end of each sample to create richer data. These processed audio files were structured into directories tagged with the accent of the speaker. 

\subsection{Separability of AccentDB: An Analysis}
Understanding the distribution of AccentDB speech recordings provides more insight into the quality of the collected data. To use the speech samples for any computational task or mathematical representation, they must first be converted to feature vectors. Mel-Frequency Cepstral Coefficient (MFCC) extraction is a very widely used technique to represent audio files as vectors. The MFCC extraction of audio clips generally produces very high-dimensional vectors (for example, \newcite{google-attention} use 40 MFCC dimensions per audio frame). We concatenated the MFCC features of each frame to obtain high-dimensional acoustic vectors for the full-length of a clip. Since modelling the distribution of high dimensional data is difficult, we performed dimensionality reduction to obtain a set of principal variables and reduce the number of random variables under consideration. Dimensionality reduction techniques, when used for speech, learn projections of high-dimensional acoustic spaces into lower dimensional spaces.
\\
The Principal Component Analysis on the acoustic vectors shows that the recordings from each accent in our collected database follows a definite convexity (Fig. \ref{fig:ex3-a}). We also performed Uniform Manifold Approximation and Projection with 20 and 50 neighbours to show that the speech samples from an accent are closer to each other (Fig. \ref{fig:ex3-b} \& Fig. \ref{fig:ex3-c}). Further, t-SNE projections of the data (Figures \ref{fig:ex3-d}, \ref{fig:ex3-e} \& \ref{fig:ex3-f}) show the separability of the accents, establishing that the speech samples collected in AccentDB model their respective accents distinctively and are well-suited for use in machine learning tasks.

\section{\label{accent_classification}Accent Classification}

\begin{table*}[ht]
\centering
\begin{tabular}{lcS[table-format=2.2]S[table-format=2.2]S[table-format=2.2]}
\toprule
\textbf{\centering Task} & \textbf{Type} & \textbf{MLP} & \textbf{CNN} & \textbf{CNN (with attention)} \\
\midrule
\textbf{Indian vs. Non Indian} & 2-class classification & {100.0\%} & {100.0\%} & {100.0\%} \\
\textbf{Non-native Indian Accents} & 4-class classification & {98.3\%} & {98.6\%} & {99.0\%} \\
\textbf{All accents} & 9-class classification & {98.4\%} & {99.3\%} & {99.5\%} \\
\bottomrule
\end{tabular}
\caption{Classification accuracy of various models on three classification tasks.}
\label{tab:classification_results}
\end{table*}


\begin{table*}[ht]
\centering


\begin{tabular*}{0.78\textwidth}{P{2.5cm}P{2.5cm}S[table-format=2.2]S[table-format=2.2]S[table-format=2.2]}
\toprule
\multicolumn{2}{c}{\textbf{Samples Used}} & \multicolumn{1}{c}{\multirow{2}{2cm}{\centering \textbf{Bangla, Telugu}}} & \multicolumn{1}{c}{\multirow{2}{2cm}{\centering \textbf{Bangla, Malayalam}}} & \multicolumn{1}{c}{\multirow{2}{2cm}{\centering \textbf{Telugu, Malayalam}}} \\
\multicolumn{1}{c}{\textbf{Training Set}} & \multicolumn{1}{c}{\textbf{Testing Set}} & \multicolumn{1}{c}{} & \multicolumn{1}{c}{} & \multicolumn{1}{c}{} \\
\midrule
\textit{$A_{1}P_{1} + A_{2}P_{1}$} & \textit{$A_{1}P_{2} + A_{2}P_{2}$} & 82.86\% & \bfseries 90.06\% & 79.75\% \\
\textit{$A_{1}P_{1} + A_{2}P_{2}$} & \textit{$A_{1}P_{2} + A_{2}P_{1}$} & 70.38\% & 74.68\% & 81.30\% \\
\textit{$A_{1}P_{2} + A_{2}P_{1}$} & \textit{$A_{1}P_{1} + A_{2}P_{2}$} & 73.35\% & 83.08\% & \bfseries 95.73\% \\
\textit{$A_{1}P_{2} + A_{2}P_{2}$} & \textit{$A_{1}P_{1} + A_{2}P_{1}$} & \bfseries 96.18\% & 76.15\% & 69.55\% \\ \bottomrule
\end{tabular*}
\caption{Model accuracy when training on speech samples from one speaker and testing on unseen samples from other speaker for  \SI{3}{} different accent pairs. \textit{A\textsubscript{\small{i}}P\textsubscript{\small{j}}} denotes all speech samples from \textit{j}-th speaker of the \textit{i}-th accent.}
\label{tab:toto}
\end{table*}

Accent classification is an important step for tasks such as speech profiling and speaker identification. The current state-of-the-art ASR systems are already within the striking range of human-level performance with word error rates (WER) as low as \SI{5.5}{\%} \cite{wer5}. Accent classification can also be used to enhance ASR systems for better generalization towards unseen data by augmenting the training dataset with more relevant features (\newcite{speech-augmention-for-asr}; \newcite{specaugment}). One such very relevant feature is present in human communication in the form of accent and hence, the task of accent classification has been crucial in the combined modelling of speech.

Over the years, multiple approaches have been used to tackle the tasks related to the accent classification for speech recognition. These include classical methods of Gaussian mixture models (GMMs) and Hidden Markov models (HMMs), machine learning models using Support Vector Machine (SVM), and very recently, deep neural architectures like Convolutional Neural Networks (CNNs) and Long Short Term Memory (LSTM).

An early architecture that was proposed for this task by \newcite{carlos-isabel-antonio} used parallel ergodic nets with context-dependent HMM units for word-level accent identification. Their system obtained a global accuracy score of \SI{65.48}{}\% on their word-level speech data comprising \SI{6}{} different accents. \newcite{ge2015accent} used purely acoustic features to build a GMM based accent classifier optimized using Heteroscedastic Linear Discriminant Analysis (HLDA). They used the FAE dataset \cite{foreign-accented-english} and achieved a success rate of \SI{51}{}\% on \SI{7}{} accents.

The recent advancements in deep learning architectures have proven to be a great success in a variety of speech recognition tasks including accent classification. In the work by \newcite{yang2018joint}, the authors highlighted the importance of accent information for acoustic modeling and presented a joint end-to-end model for multi-accent speech recognition that achieves significant improvement on word-error rates. They used a bi-directional LSTM model with average pooling, and trained it with a Connectionist Temporal Classification (CTC) loss function. \newcite{bird2019accent} explored a variety of different techniques for accent classification on diphthong vowel sounds collected from speakers from Mexico and the United Kingdom. They achieved a classification accuracy of \SI{94.74}{}\% using an ensemble model of Random Forest and LSTM.

\subsection{\label{experiments}Experiments}
We ran classification experiments on our database using two standard baseline neural network architectures - a multi layer perceptron (MLP) and a CNN model. We evaluated the classification models in three different setups - (i) classifying
amongst Indian accents collected in our database and non-Indian accents obtained from AWS Polly, (ii) classifying amongst the 4 collected Indian accents in our database, (iii) and finally classifying amongst all the 9 accents in AccentDB.

\subsubsection{\label{preprocessing} Preprocessing}
Each audio file was divided into 10ms segments with a 1ms overlap between the segments. All the samples were less than \SI{5}{} seconds in duration and hence padded to a standardized input dimension of \SI{499}{}. For each of these segments, we extracted \SI{13}{} MFCC features. Hence, our final vector input for \textit{n} audio files is of the dimension (n, \SI{}{499}, \SI{13}{}). This two-dimensional image-like vector for each audio file was used as the input to the first convolutional layer in all the CNN-based models. For MLP models, the input vector was created by flattening the image to one dimension.

\subsubsection{Model Architecture and Training}
The MLP model consists of multiple fully connected layers stacked together. The CNN model uses a combination of 1D Convolutional and Max Pooling layers, followed by multiple dense fully connected layers. For calculating the class probabilities, softmax activation was used in the final layer of each model. We used Adam \cite{adam} and RMSProp \cite{rmsprop} optimizers, with a learning rate of \SI{0.001}{} using cross-entropy loss function. Dropout was used in dense layers for regularization. A variety of batch sizes were tried during training to achieve the best results. As part of the evaluation, we used \SI{20}{}\% of the total data present as test set for evaluation. 
\\
As the next step, we augmented the CNN network with attention. Attention mechanism has been successfully applied in machine translation \cite{attentionMT} and image captioning \cite{showAttendTell}. Promising results have also been obtained for speech based tasks, e.g., in \cite{attentionSpeech2015}, where the authors solved the task of acoustic scene classification using a Convolutional, Long Short-term memory, Deep Neural Network (CLDNN) network and several attention-based LSTM models. We took this motivation further to apply attention mechanism onto the accent data to analyze the segments of the audio that are given more importance by our classification model. We used multiple variations of attention, firstly \SI{1}{}D and \SI{2}{}D variations based on the number of dimensions used. In the \SI{1}{}D version, attention vector is shared across the input dimensions, which correspond to the number of MFCC features used (\SI{13}{} in our case). For the \SI{2}{}D version, separate attention probability vectors were learnt for each input feature dimension. We also varied the layer to the output of which attention is applied.

\subsection{Results}
We evaluate our MLP, CNN and attention-CNN models on three different classification tasks, as described in the section \ref{experiments} The accuracy results are summarized in Table \ref{tab:classification_results}. As is observable in the results, all the models performed exceptionally well, with the CNN models having a slight edge in accuracy as expected. Particularly in the binary classification setup, these models were able to detect the correct class with \SI{100}{}\% accuracy. These instances of high accuracy can also be attributed to the presence of a quality dataset with good separability as discussed in section \SI{2.6}{}.

\begin{figure}[h]
\begin{center}
\centering
\includegraphics[width=0.45\textwidth]{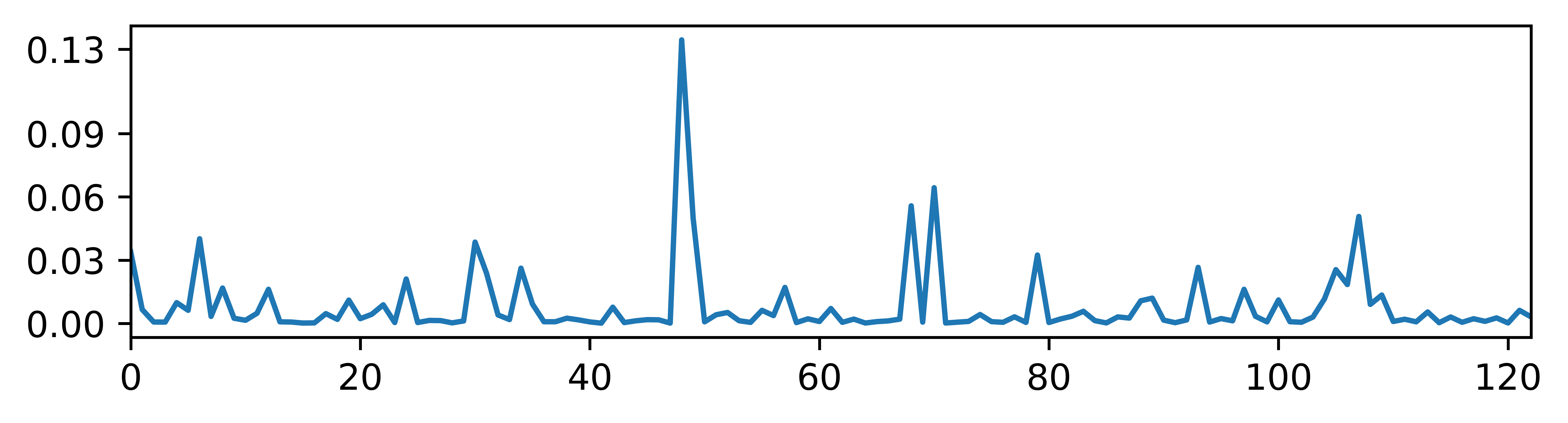}
\includegraphics[width=0.45\textwidth]{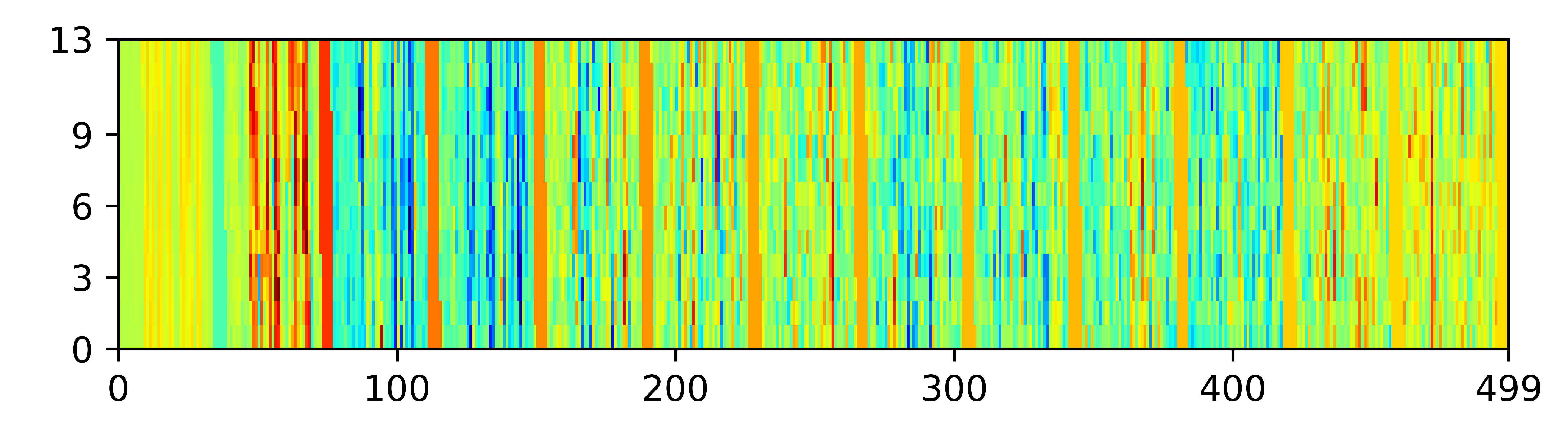}
\caption{Time aligned attention scores with MFCC features corresponding to the sentence: \textit{"Four hours of steady work faced us."}}
\label{attention-MFCC-1}
\end{center}
\end{figure}

\begin{figure}[h]
\begin{center}
\centering
\includegraphics[width=0.45\textwidth]{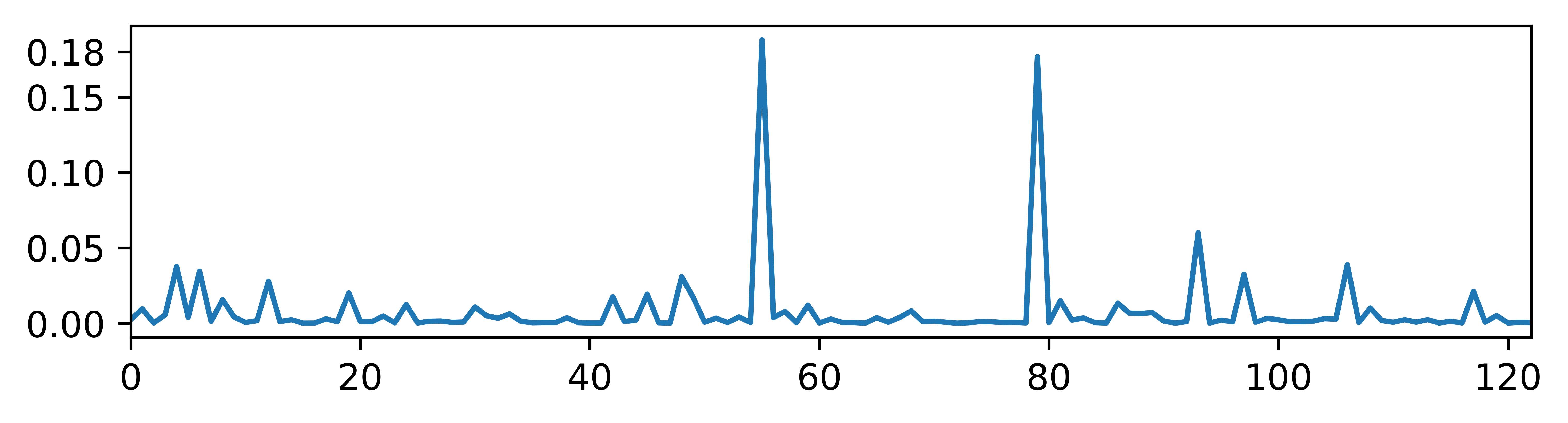}
\includegraphics[width=0.45\textwidth]{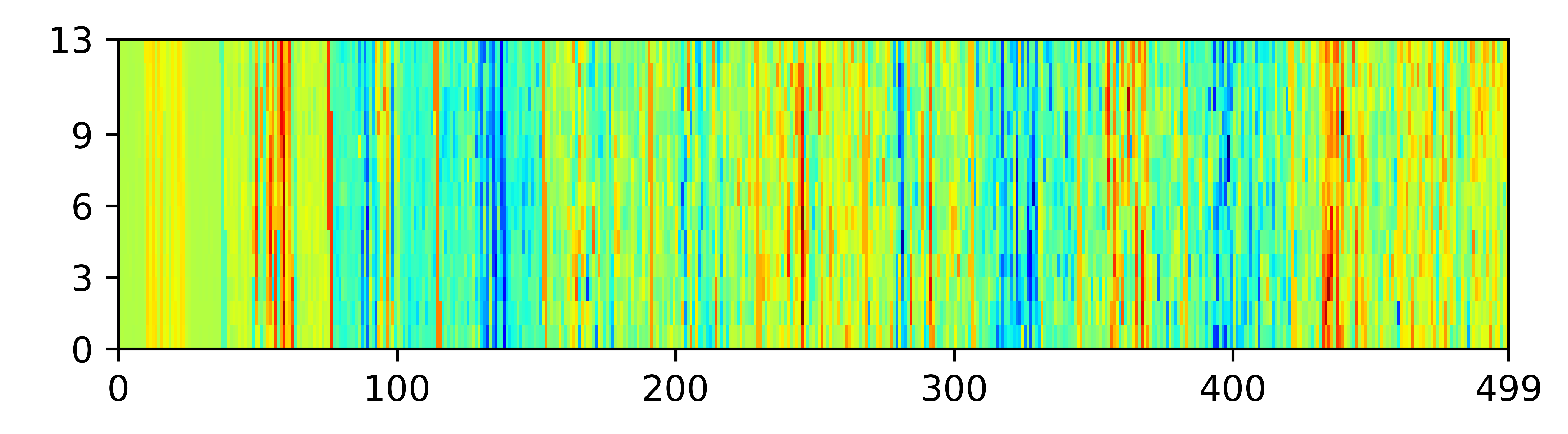}
\caption{Time aligned attention scores with MFCC features corresponding to the sentence: \textit{"It's easy to tell the depth of a well."}}
\label{attention-MFCC-2}
\end{center}
\end{figure}

\subsection{Train on One, Test on Other}
Speech classification models tend to overfit if they have a large number of trainable parameters but the training data is not extensive enough. This leads to poor generalization of models from training samples to unseen samples. To test if the models described previously, perform well even on unseen data, we evaluated our models in a challenging setup. Three accents in AccentDB - Bangla, Telugu and Malayalam were chosen for this experiment and the data for each accent was split into two, based on the speaker. We trained binary classifier models on sets of two accents by feeding them with only one half of the data of each accent (i.e.\ data of one speaker per accent). The models were tested on the unseen half data of each accent (i.e.\ the other speaker). Table \ref{tab:toto} shows that our classifier models generalized well on the test data consisting of samples from other speakers even without seeing them during training.



\subsection{Interpreting Attention}

The attention scores that were obtained were analyzed by plotting them against the corresponding MFCC features for two audio files of Malayalam accent. In Figure \ref{attention-MFCC-1}, we observe a clear spike around the word \textit{"Four"}, while in figure \ref{attention-MFCC-2}, the spikes correspond to timestamps around the words \textit{"depth"} and \textit{"well"}. 
These can be attributed to the different pronunciations of a particular phoneme sequence. For example, "depth" has the sounds that don't occur next to each other in the phonetics of Indian languages. So, each participant looks up to their own phonology to pronounce the word.

\section{Accent Neutralization}
State-of-the-art ASR systems often do not perform well on rare non-native accents, primarily due to the non-availability of good quality data for training such systems. We present our dataset on Indian Accents to augment training data for existing ASR systems to help make them more robust. ASR systems that perform well on native accents can further be improved for rare accents by performing accent neutralization. This means processing non-native audio file to make it sound like that of a native accent that the ASR system performs well on. The accent neutralization performed here involves extracting and transforming the para-linguistic and non-linguistic features of a source accent into those of a target accent while preserving the linguistic features. 
Acoustic feature conversions have been explored in other speech processing tasks as well. \newcite{voice-conversion-challenge} devised a challenge to better understand transformations of voice identity among speakers. For accent conversion, \newcite{foreign-english-accent-conversion} proposed a method to create accented samples of words by leveraging the difference between a dialect and General American English. Their model learns generalizations that would otherwise be created using rules written manually by phonologists. 

With the success of neural networks in speech modelling, recent works have attempted end-to-end accented speech recognition. The experiments performed by \newcite{bearman-accent-stanford} and \newcite{end-to-end-speech} are on datasets that consist primarily of native accents (such as American, British, Australian, Canadian) and Indian as well, but the performance of models are underutilized due to the absence of non-native accents data. As reported in the next sections, we utilized the data collected in AccentDB to train and test deep neural networks on the task of non-native accent neutralization. We propose these transformation models to be used as an inference-time pre-processing step for ASR systems in order to overcome challenges associated with low resource accents.

\begin{figure}[hb]
\begin{center}
\includegraphics[width=0.40\textwidth]{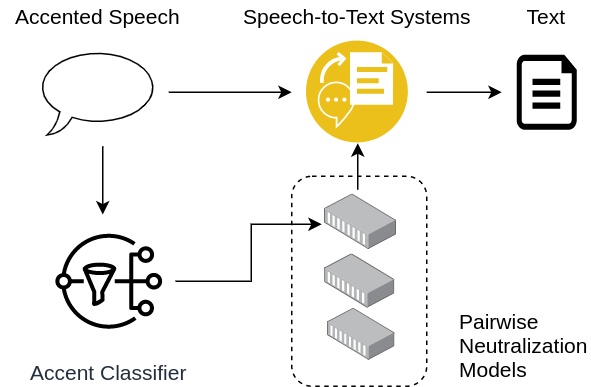} 
\caption{Use of pairwise accent neutralization in speech-to-text systems.}
\label{accent-neutralization}
\end{center}
\end{figure}

\subsection{Pairwise Neutralization}
A pairwise accent neutralization system consists of a set of individual models which can convert MFCC feature vectors of samples belonging to a source accent to those of samples belonging to a target accent. This set of individual converter systems can be used in conjunction with an accent classification system. An input audio file is routed to the converter corresponding to its predicted accent class from the classifier. The selected converter would be the one which can convert files belonging to this predicted accent to the given target accent. (Figure \ref{accent-neutralization}). The pre-processing step for this experiment is the same as that described in section \ref{preprocessing}

\subsubsection{\label{autoenc}Model Architecture}
 We trained a stacked denoising autoencoder network \cite{stackeddae} consisting of a series of convolution and pooling layers followed by deconvolutional (deconvolutional layers are required for upsampling) and pooling layers. The output of each layer is passed through a \textit{tanh} activation function. Further, we train another similar network for evaluating reconstruction on reversed pairs where the source and target files are swapped. The autoencoder network's loss function is defined by feature-wise mean squared error between the input and output vectors. We used RMSProp optimizer \cite{rmsprop} with a learning rate of \SI{0.001}{}. The convolutional layers act as feature extractors for the input MFCC feature vectors and learn to encode them into a dense representation. The deconvolutional layers learn transformations on this dense representation for reconstruction into MFCC features of the target accent.

\begin{table}[h]
\centering
\begin{tabular}{llS[table-format=2.2]S[table-format=2.2]}
\toprule
\multicolumn{1}{p{1.3cm}}{\centering \textbf{Source\\ Accent}} &
\multicolumn{1}{p{1.3cm}}{\centering \textbf{Target\\ Accent}} &
\multicolumn{1}{p{1.5cm}}{\centering \textbf{Accuracy}} &
\multicolumn{1}{p{1.5cm}}{\centering \textbf{Accuracy\\(reverse)}}
\\
\midrule
\rule{0mm}{2mm}
\multirow{4}{*}{Bangla}    & American  &  \bfseries  99.21\%      &     98.67\%                      \\
                        & Australian    &    99.02\%    &       85.52\%                    \\
                        & British          &  95.17\%        &     97.36\%            \\
                        & Welsh         &   98.96\%       &  \bfseries 98.73\%     \\          \midrule
\rule{0mm}{2mm}

\multirow{4}{*}{Indian} & American  & \bfseries 98.77\%     &    95.23\%          \\
                        &  Australian      &    98.63\%      &         \bfseries 95.77\%        \\
                        &  British           &   95.27\%       &     92.64\%            \\
                        & Welsh         &      97.27\%    &  90.48\%         \\       
                        \midrule
\rule{0mm}{2mm}
Odiya                & Malayalam  &   \bfseries   87.84\%    &     \bfseries  93.11\%        \\  \bottomrule
\end{tabular}
\caption{Classification accuracy on pairwise neutralization of \SI{9}{} accent pairs.}
\label{tab:pairwise_neutralization}
\end{table}

\subsubsection{Results}
The reconstructed feature vectors obtained from the autoencoder model were evaluated on classification accuracy metric using CNN classifiers trained in section \ref{accent_classification} We performed this experiment for neutralizing Bangla and Indian accents into 4 native accents. Our model performed very well on the non-native to native accent neutralization achieving an accuracy of \textgreater \SI{95}{}\% on all \SI{8}{} experiments. We obtained an accuracy of \textgreater \SI{85}{}\% when converting from native to non-native accents as well. The model can also be used to neutralize a non-native accent into a different non-native accent as shown through the Odiya-Malayalam pair. The results in Table \ref{tab:pairwise_neutralization} show that the  transformations learnt through our model can be used effectively as a preprocessing step in ASR systems enabling them to work well on non-native accents.

\begin{table}[ht]
\centering
\begin{tabular}{m{3.3cm}m{1.4cm}m{1.4cm}}
\toprule
\textbf{Model} & \textbf{1 source 2\,\, targets} & \textbf{2 sources 2\, targets} \\
\midrule
CNN Autoencoder + Skip\, Connections &  {52.15\%} &  {66.73\%} \\ \hline
LSTM Autoencoder + Skip\, Connections &  {53.46\%} &  {70.08\%} \\ 
\bottomrule
\end{tabular}
\caption{Classification accuracy on multi-source multi-target accent neutralization.}
\label{tab:msmt}
\end{table}

\subsection{Multi-source Accent Multi-target Accent Neutralization}
Extending the neutralization task for a set of \textit{n} accents requires training $2 \times \binom n2$  pairwise neutralization models. Moreover, any device using this system would also require the source accent to be identified first before choosing a pairwise trained model to perform neutralization. To overcome both of these challenges, we present a single model that can be trained over multiple accents to neutralize samples from  \textit{S\textsubscript{n}} number of source accents to \textit{T\textsubscript{n}} number of target accents.

\subsubsection{Preprocessing}
To train a single model with pairs of \textit{(source, target)} samples belonging to multiple accents, we added an additional marker in each training pair, similar to zero-shot neural machine translation system proposed in \cite{googlezeroshotnmt}. The MFCC feature vectors of source accent samples were prefixed with a  \SI{13}{}-dimensional one-hot encoded representation of accent label of target samples. Hence, the transformation of each input vector of dimension {\small{(\SI{499}{},  \SI{13}{})}} is as follows:
\[{\textit{S\textsubscript{\large{i}}}\textsuperscript{\textit{{(499, 13)}}} \rightarrow \textit{T\textsubscript{\large{j}}}}\textsuperscript{\textit{
{(499, 13)}}} 
\Rightarrow
\textit{(L\textsubscript{\small{T\textsubscript{\small{j}}}} . S\textsubscript{\large{i}})}\textsuperscript{\textit{{(500, 13)}}} \rightarrow \textit{T\textsubscript{\large{j}}}\textsuperscript{\textit{{(499, 13)}}}
\]

where \textit{S\textsubscript{\small{i}}} denotes input files of \textit{i-th} source accent, \textit{T\textsubscript{\small{j}}} denotes target files of the \textit{j-th} accent, \textit{L\textsubscript{\small{T\textsubscript{\small{j}}}}} denotes label of \textit{T\textsubscript{\small{j}}-th} target accent and (.) represents concatenate operation.

\subsubsection{Experiments and Results}
We used the prefixed inputs to run experiments in two setups for this task. We started with a set of \SI{3}{} accents such that all the samples wee from same source accent and wee to be neutralized into two different target accents. We then experimented with the same set of \SI{3}{} accents but now with samples from two different source accents.
The convolutional autoencoder described in section \ref{autoenc} was augmented with skip connections to propagate target accent information in the form of label vector. This target accent label information is available in each layer up until the last one. We also experimented with a stacked LSTM autoencoder with skip connections. Table \ref{tab:msmt} compiles our preliminary results.

\section{Conclusion and Future Works}
We presented AccentDB, a well-labelled parallel database of non-native accents that shall aid in the development of machine learning models for speech recognition. Having a parallel corpus is better suited for tasks such as accent neutralization where each source sample should correspond to a target sample with the same vocabulary such that the differences in accent could be  modelled easily. We evaluated accent classification models in a variety of settings and also discussed an interpretation of attention scores for analyzing audio frames. Finally, we showed the applicability of autoencoder models for accent neutralization. Future scope of our work includes enriching the database with more accents, and a larger variety of speakers in terms of age and gender. We would also like to add single-word database, ideally labelled for phonemes to have the data devoid of the effects of suprasegmental features. 

\section{Acknowledgements}

This study was funded by the OPERA grant from BITS Pilani, provided to Dr. Pranesh Bhargava.

\section{Bibliographical References}
\label{main:ref}

\bibliographystyle{lrec}
\bibliography{lrec2020W-xample}

\begin{thebibliography}{}

\bibitem[\protect\citename{Bahdanau \bgroup et al.\egroup }2014]{attentionMT}
Bahdanau, D., Cho, K., and Bengio, Y.
\newblock (2014).
\newblock Neural machine translation by jointly learning to align and
  translate.
\newblock {\em arXiv preprint arXiv:1409.0473}.

\bibitem[\protect\citename{Bearman \bgroup et al.\egroup
  }2017]{bearman-accent-stanford}
Bearman, A., Josund, K., and Fiore, G.
\newblock (2017).
\newblock Accent conversion using artificial neural networks.

\bibitem[\protect\citename{Bird \bgroup et al.\egroup }2019]{bird2019accent}
Bird, J.~J., Wanner, E., Ek{\'a}rt, A., and Faria, D.~R.
\newblock (2019).
\newblock Accent classification in human speech biometrics for native and
  non-native english speakers.

\bibitem[\protect\citename{Chorowski \bgroup et al.\egroup
  }2015]{attentionSpeech2015}
Chorowski, J.~K., Bahdanau, D., Serdyuk, D., Cho, K., and Bengio, Y.
\newblock (2015).
\newblock Attention-based models for speech recognition.
\newblock In {\em Advances in neural information processing systems}, pages
  577--585.

\bibitem[\protect\citename{Ge \bgroup et al.\egroup }2015]{ge2015accent}
Ge, Z., Tan, Y., and Ganapathiraju, A.
\newblock (2015).
\newblock Accent classification with phonetic vowel representation.
\newblock In {\em 2015 3rd IAPR Asian Conference on Pattern Recognition
  (ACPR)}, pages 529--533. IEEE.

\bibitem[\protect\citename{Gemmeke \bgroup et al.\egroup }2017]{audioset}
Gemmeke, J.~F., Ellis, D. P.~W., Freedman, D., Jansen, A., Lawrence, W., Moore,
  R.~C., Plakal, M., and Ritter, M.
\newblock (2017).
\newblock Audio set: An ontology and human-labeled dataset for audio events.
\newblock In {\em Proc. IEEE ICASSP 2017}, New Orleans, LA.

\bibitem[\protect\citename{Guo \bgroup et al.\egroup }2017]{google-attention}
Guo, J., Xu, N., Li, L.-J., and Alwan, A.
\newblock (2017).
\newblock Attention based cldnns for short-duration acoustic scene
  classification.

\bibitem[\protect\citename{Hernandez \bgroup et al.\egroup
  }2018]{deep-learning-for-video-games}
Hernandez, S.~P., Bulitko, V., Carleton, S., Ensslin, A., and Goorimoorthee, T.
\newblock (2018).
\newblock Deep learning for classification of speech accents in video games.
\newblock In {\em Joint Proceedings of the {AIIDE} 2018 Workshops 2018}.

\bibitem[\protect\citename{IEEE}1969]{harvard-sentences}
IEEE.
\newblock (1969).
\newblock Ieee recommended practice for speech quality measurements.
\newblock {\em IEEE No 297-1969}, pages 1--24, June.

\bibitem[\protect\citename{Johnson \bgroup et al.\egroup
  }2017]{googlezeroshotnmt}
Johnson, M., Schuster, M., Le, Q.~V., Krikun, M., Wu, Y., Chen, Z., Thorat, N.,
  Vi{\'e}gas, F., Wattenberg, M., Corrado, G., et~al.
\newblock (2017).
\newblock Google's multilingual neural machine translation system: Enabling
  zero-shot translation.
\newblock {\em Transactions of the Association for Computational Linguistics},
  5:339--351.

\bibitem[\protect\citename{Kingma and Ba}2014]{adam}
Kingma, D.~P. and Ba, J.
\newblock (2014).
\newblock Adam: A method for stochastic optimization.
\newblock {\em arXiv preprint arXiv:1412.6980}.

\bibitem[\protect\citename{Kitashov \bgroup et al.\egroup
  }2018]{foreign-english-accent-conversion}
Kitashov, F., Svitanko, E., and Dutta, D.
\newblock (2018).
\newblock Foreign english accent adjustment by learning phonetic patterns.
\newblock {\em CoRR}, abs/1807.03625.

\bibitem[\protect\citename{Ko \bgroup et al.\egroup
  }2015]{speech-augmention-for-asr}
Ko, T., Peddinti, V., Povey, D., and Khudanpur, S.
\newblock (2015).
\newblock Audio augmentation for speech recognition.
\newblock In {\em INTERSPEECH}, pages 3586--3589. ISCA.

\bibitem[\protect\citename{Kominek and Black}2004]{cmu-arctic}
Kominek, J. and Black, A.
\newblock (2004).
\newblock The cmu arctic speech databases.
\newblock {\em SSW5-2004}, 01.

\bibitem[\protect\citename{Ladefoged}1993]{Ladefoged}
Ladefoged, P.
\newblock (1993.).
\newblock {\em A Course in phonetics.}
\newblock Harcourt,, Firt Worth :, 3rd ed. edition.

\bibitem[\protect\citename{Lander}2007]{foreign-accented-english}
Lander, T.
\newblock (2007).
\newblock Cslu foreign accented english release 1.2 ldc2007s08.

\bibitem[\protect\citename{Mazzoni}1999]{audacity}
Mazzoni, D.
\newblock (1999).
\newblock Audacity \textregistered software is copyright \textcopyright
  1999-2019 audacity team.

\bibitem[\protect\citename{Nagrani \bgroup et al.\egroup }2017]{voxceleb}
Nagrani, A., Chung, J.~S., and Zisserman, A.
\newblock (2017).
\newblock Voxceleb: a large-scale speaker identification dataset.
\newblock {\em CoRR}.

\bibitem[\protect\citename{Park \bgroup et al.\egroup }2019]{specaugment}
Park, D.~S., Chan, W., Zhang, Y., Chiu, C.-C., Zoph, B., Cubuk, E.~D., and Le,
  Q.~V.
\newblock (2019).
\newblock Specaugment: A simple data augmentation method for automatic speech
  recognition.
\newblock {\em arXiv preprint arXiv:1904.08779}.

\bibitem[\protect\citename{Saon \bgroup et al.\egroup }2017]{wer5}
Saon, G., Kurata, G., Sercu, T., Audhkhasi, K., Thomas, S., Dimitriadis, D.,
  Cui, X., Ramabhadran, B., Picheny, M., Lim, L., Roomi, B., and Hall, P.
\newblock (2017).
\newblock English conversational telephone speech recognition by humans and
  machines.
\newblock {\em CoRR}, abs/1703.02136.

\bibitem[\protect\citename{Teixeira \bgroup et al.\egroup
  }1996]{carlos-isabel-antonio}
Teixeira, C., Trancoso, I., and Serralheiro, A.
\newblock (1996).
\newblock Accent identification.
\newblock 01.

\bibitem[\protect\citename{Tieleman and Hinton}2012]{rmsprop}
Tieleman, T. and Hinton, G.
\newblock (2012).
\newblock {Lecture 6.5---RmsProp: Divide the gradient by a running average of
  its recent magnitude}.
\newblock COURSERA: Neural Networks for Machine Learning.

\bibitem[\protect\citename{Toda \bgroup et al.\egroup
  }2016]{voice-conversion-challenge}
Toda, T., Chen, L.-H., Saito, D., Villavicencio, F., Wester, M., Wu, Z., and
  Yamagishi, J.
\newblock (2016).
\newblock The voice conversion challenge 2016.

\bibitem[\protect\citename{Viglino \bgroup et al.\egroup
  }2019]{end-to-end-speech}
Viglino, T., Motlicek, P., and Cernak, M.
\newblock (2019).
\newblock End-to-end accented speech recognition.
\newblock {\em Proc. Interspeech 2019}, pages 2140--2144.

\bibitem[\protect\citename{Vincent \bgroup et al.\egroup }2010]{stackeddae}
Vincent, P., Larochelle, H., Lajoie, I., Bengio, Y., and Manzagol, P.-A.
\newblock (2010).
\newblock Stacked denoising autoencoders: Learning useful representations in a
  deep network with a local denoising criterion.
\newblock {\em J. Mach. Learn. Res.}, 11:3371--3408, December.

\bibitem[\protect\citename{Weinberger and Kunath}2011]{please-call-stella}
Weinberger, S. and Kunath, S.
\newblock (2011).
\newblock The speech accent archive: Towards a typology of english accents.
\newblock {\em Language and Computers}, 73, 12.

\bibitem[\protect\citename{Xu \bgroup et al.\egroup }2015]{showAttendTell}
Xu, K., Ba, J., Kiros, R., Cho, K., Courville, A., Salakhudinov, R., Zemel, R.,
  and Bengio, Y.
\newblock (2015).
\newblock Show, attend and tell: Neural image caption generation with visual
  attention.
\newblock In {\em International conference on machine learning}, pages
  2048--2057.

\bibitem[\protect\citename{Yang \bgroup et al.\egroup }2018]{yang2018joint}
Yang, X., Audhkhasi, K., Rosenberg, A., Thomas, S., Ramabhadran, B., and
  Hasegawa-Johnson, M.
\newblock (2018).
\newblock Joint modeling of accents and acoustics for multi-accent speech
  recognition.
\newblock In {\em 2018 IEEE International Conference on Acoustics, Speech and
  Signal Processing (ICASSP)}, pages 1--5. IEEE.

\end{thebibliography}


\end{document}